\documentclass[preprint,nofootinbib,floats,superscriptaddress,tightenlines,amsfonts,amsmath,amssymb]{revtex4}
\pdfoutput=1

\usepackage{graphicx}
\usepackage{bm}
\usepackage{color}
\usepackage{multirow}
\usepackage{slashed}
\usepackage[colorlinks,linkcolor=black,anchorcolor=black,citecolor=black]{hyperref}

\begin{document}
\baselineskip=17pt \parskip=7pt

\preprint{NCTS-PH/1726}
\hspace*{\fill}

\title{Connecting $\bm{b\to s\ell\bar\ell}$ anomalies to enhanced rare nonleptonic
$\bm{\bar{B}_s^0}$ decays in $\bm{Z'}$ model}

\author{Gaber Faisel}
\email{gaberfaisel@sdu.edu.tr}
\affiliation{Department of Physics, Faculty of Arts and Sciences,
S\"uleyman Demirel University, D685, Isparta 32260, Turkey}

\author{Jusak Tandean}
\email{jtandean@yahoo.com}
\affiliation{Department of Physics, National Taiwan University, \\
No.\,\,1, Sec.\,\,4, Roosevelt Rd., Taipei 10617, Taiwan}
\affiliation{Physics Division, National Center for Theoretical Sciences, \\
No.\,\,101, Sec.\,\,2, Kuang Fu Rd., Hsinchu 30013, Taiwan\bigskip}


\begin{abstract}

The present data on a number of observables in $b\to s\mu^+\mu^-$ processes manifest some
tensions with the standard model (SM).
Assuming that these anomalies have a new physics origin, we consider the possibility that a $Z'$
boson is responsible for them.
We further assume that its interactions with quarks also affect rare nonleptonic decays of
the $\bar B_s^0$ meson which are purely isospin-violating and tend to be dominated by
electroweak-penguin contributions, namely $\bar B_s^0\to(\eta,\eta',\phi)(\pi^0,\rho^0)$.
Most of these decays are not yet observed, and their rates are expected to be relatively small
in the SM.
Taking into account constraints from various measurements, including the evidence for
$\bar B_s^0\to\phi\rho^0$ recently seen by LHCb, we find that the $Z'$ effects on
$\bar B_s^0\to(\eta,\phi)\pi^0$ can make their rates bigger than the SM predictions
by up to an order of magnitude.
For $\bar B_s^0\to\eta'\pi^0,(\eta,\eta')\rho^0$, the enhancement factors are at most a few.
Since the $Z'$ contributions to the different channels depend on different combinations of its
couplings, observations of more of these decays in future experiments, along with improved
$b\to s\mu^+\mu^-$ data, will probe this $Z'$ scenario more thoroughly.

\end{abstract}

\maketitle

\section{Introduction \label{sec:intro}}

The latest measurements of various \,$b\to s\mu^+\mu^-$\, processes have turned up some
intriguing discrepancies from the expectations of the standard model (SM) of particle physics.
Specifically, the LHCb Collaboration in its angular analysis of the decay
$B^0\to K^{*0}\mu^+\mu^-$ found tensions with the SM at the $3.4\sigma$
level\,\,\cite{Aaij:2015oid}.
This was later confirmed in the Belle experiment on the same process, but with lower
statistical confidence\,\,\cite{Wehle:2016yoi}.
Furthermore, LHCb findings\,\,\cite{Aaij:2014ora,Aaij:2017vbb} on the ratio $R_K$ of
the branching fractions of $B^+\to K^+\mu^+\mu^-$ and $B^+\to K^+e^+e^-$ decays and on
the corresponding ratio $R_{K^*}$ for $B^0\to K^{*0}\mu^+\mu^-$ and $B^0\to K^{*0}e^+e^-$
decays are all below their SM predictions\,\,\cite{Hiller:2003js,Bouchard:2013mia,Bordone:2016gaq}
by 2.1$\sigma$ to 2.6$\sigma$.
In addition, the current data\,\,\cite{Aaij:2014pli,Aaij:2015esa,pdg} on the branching
fractions of $B\to K^{(*)}\mu^+\mu^-$ and $B_s\to\phi\mu^+\mu^-$ favor values less than their
SM estimates.

Although the statistical significance of the aforesaid deviations from SM expectations is
still too low for a definite conclusion, they may be early clues about interactions beyond
the SM in $b\to s$ transitions.
Recent model-independent theoretical analyses have in fact demonstrated that new physics (NP)
could account for these anomalies\,\,\cite{Descotes-Genon:2013wba,Beaujean:2013soa,
Hurth:2013ssa,Alonso:2014csa,Hiller:2014yaa,Ghosh:2014awa,Hurth:2014vma,Capdevila:2017bsm,
Altmannshofer:2017yso,DAmico:2017mtc,Hiller:2017bzc,Geng:2017svp,Ciuchini:2017mik,
Celis:2017doq,Hurth:2017hxg}.
In view of the possibility that these tentative hints of NP will be confirmed by upcoming
experiments, it is of interest to explore the potential implications for other \,$b\to s$\,
processes.

Among them are the nonleptonic decays \,$\bar B_s\to\eta\pi^0$, \,$\bar B_s\to\eta'\pi^0$,
\,$\bar B_s\to\phi\pi^0$, \,$\bar B_s\to\eta\rho^0$, \,$\bar B_s\to\eta'\rho^0$,\, and
\,$\bar B_s\to\phi\rho^0$.\,
Each of these transitions has a final state with total isospin \,$I=1$\, and thus
fully breaks isospin symmetry, implying that their amplitudes receive no
contributions from QCD-penguin operators and arise instead from charmless tree and
electroweak-penguin (EWP) operators\,\,\cite{Fleischer:1994rs,Deshpande:1994yd}.
The product of Cabibbo-Kobayashi-Maskawa (CKM) matrix elements in the tree contributions is
suppressed compared to that in the EWP ones, and the suppression factor is
\,$|V_{us}V_{ub}|/|V_{ts}V_{tb}|\sim0.02$.
Consequently, although the Wilson coefficients of the tree operators are much bigger than
those of the EWP operators, the latter turn out to dominate
the majority of these channels, and the resulting decay rates are relatively
low\,\,\cite{Fleischer:1994rs,Deshpande:1994yd, Tseng:1998wm}.
Most of them are not yet observed, the exception being $\bar B_s\to\phi\rho^0$.
Evidence for it was detected by LHCb last year\,\,\cite{Aaij:2016qnm} with a\,\,branching
fraction
\,${\cal B}\big(\bar B_s\to\phi\rho^0\big)=(0.27\pm0.08)\!\times\!10^{-6}$ \cite{pdg},
which agrees with some of its estimates in the SM within sizable
errors\,\,\cite{Ali:2007ff,Cheng:2009mu,Wang:2017rmh,Faisel:2011kq,Hofer:2010ee}.

The smallness of the rates of \,$\bar B_s\to(\eta,\eta',\phi)(\pi^0,\rho^0)$\, in
the SM implies that they may serve as probes of physics beyond it.
This has been considered to varying extents in the contexts of different
models\,\,\cite{Faisel:2011kq,Hofer:2010ee,Hua:2010wf,Chang:2013hba,Faisel:2014dna,Bobeth:2014rra}.
In this paper, we treat these rare nonleptonic $\bar B_s$ decays along similar lines and suppose
that the NP influencing them also causes the aforementioned $b\to s\mu^+\mu^-$ anomalies.
We adopt in particular a scenario where an electrically neutral and uncolored spin-one particle,
the $Z'$ boson, is responsible for the new interactions in these two sets of \,$b\to s$\,
transitions.
We assume that it couples nonuniversally to SM fermions and does not mix with SM gauge bosons,
but it is not necessarily a gauge boson and could even be composite.

Although the possibility of NP effects on \,$\bar B_s\to(\eta,\eta',\phi)(\pi^0,\rho^0)$\,
in the $Z'$ context has been entertained before \cite{Hofer:2010ee,Hua:2010wf,Chang:2013hba,
Faisel:2014dna}, our current paper contains new considerations and results which were not
available in the previous literature.
Firstly, while the past studies separately examined only subsets of these six modes,\footnote{Of
the six modes, only \,$\bar B_s\to\phi\pi^0$\, was discussed in \cite{Hua:2010wf,Chang:2013hba},
\,$\bar B_s\to\phi(\pi^0,\rho^0)$\, in \cite{Hofer:2010ee}, and
\,$\bar B_s\to(\eta,\eta')(\pi^0,\rho^0)$\, in \cite{Faisel:2014dna}.}
here we treat all of them at the same time, noting that among $B_s$ decays into two charmless
mesons the six are the only ones which are strangeness changing and purely isospin-violating.
This allows us to gain a more complete picture than before concerning the $Z'$ contributions,
which reveals clearly how they in general modify the different channels in different ways.
A second novel aspect of our analysis is that, as stated in the preceding paragraph,
we explore a scenario in which the same $Z'$ not only modifies these nonleptonic $\bar B_s$
decays, but also gives rise to the \,$b\to s\ell\bar\ell$\, anomalies.
It turns out that assuming this link between the two sets of processes leads to an important
consequence for the $Z'$ interactions, namely that the left-handed $bsZ'$ coupling must be
roughly ten times stronger than the right-handed one if both of them exist,
as will be detailed later on.
This particular finding was absent from the earlier studies\,\,\cite{Hua:2010wf,
Chang:2013hba,Hofer:2010ee,Faisel:2014dna}, which did not deal with such a potential link,
as most of them appeared before the arrival of the anomalies.
A third significant novelty in our analysis is that we take into account the foregoing
evidence of \,$\bar B_s\to\phi\rho^0$\, recently seen by LHCb\,\,\cite{Aaij:2016qnm}.
As this new measurement, albeit still with a sizable uncertainty, is compatible
with its SM expectations, we will show that the implied room for the $Z'$ influence on
the channels with the $\rho^0$, not only \,$\bar B_s\to\phi\rho^0$\,
but also \,$\bar B_s\to(\eta,\eta')\rho^0$,\, is now limited.
In contrast, previously \,$\bar B_s\to\phi\rho^0$\, and \,$\bar B_s\to\eta\rho^0$\,
were allowed in Refs.\,\,\cite{Hofer:2010ee} and \cite{Faisel:2014dna}, respectively,
to have rates exceeding their SM predictions by an order of magnitude.
Last but not least, we will nevertheless also demonstrate that the viable $Z'$ parameter
space still accommodates the possibility that the pionic channels \,$\bar B_s\to\phi\pi^0$\,
and \,$\bar B_s\to\eta\pi^0$\, can have rates which are about a factor of ten higher than
their SM values.
Needless to say, this should add to the motivation for intensified efforts in upcoming
experiments at LHCb and Belle II to investigate all these decays.
The acquired data on them would provide especially useful complementary information about
the NP responsible for the \,$b\to s\ell\bar\ell$\, anomalies should the latter be
established by future measurements to be signals of physics beyond the SM.

The rest of the paper is organized as follows.
In Sec.\,\,\ref{ffz'}, we address the $Z'$ contributions to \,$b\to s\mu^+\mu^-$\, and apply
constraints from the relevant empirical information, including that on
$B_s$-$\bar B_s$ mixing.
In Sec.\,\,\ref{Bs2MM'}, we examine the impact of the $Z'$ interactions with SM quarks on
\,$\bar B_s\to(\eta,\eta',\phi)(\pi^0,\rho^0)$.\,
To evaluate their amplitudes, we employ the soft-collinear effective theory\,\,\cite{Bauer:2000ew,
Bauer:2000yr,Chay:2003zp,Bauer:2004tj,Bauer:2005kd,Williamson:2006hb,Wang:2008rk}.
We show that in the $Z'$ presence the rates of \,$\bar B_s\to\eta'\pi^0,(\eta,\eta')\rho^0$\, can
increase by as much as factors of a few with respect to the predictions in the SM, while the rates
of \,$\bar B_s\to(\eta,\phi)\pi^0$\, can exceed their SM values by up to an order of magnitude.
We make our conclusions in Sec.\,\,\ref{conclusions}.

\section{\boldmath$Z'$ interactions in $b\to s\mu\bar\mu$ and $B_s$-$\bar B_s$ mixing\label{ffz'}}

\begin{figure}[b] \bigskip
\includegraphics[width=5cm]{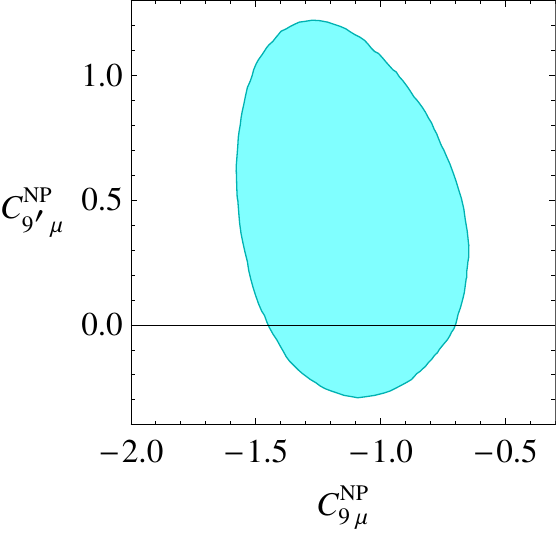}\vspace{-3pt}
\caption{Allowed 2$\sigma$ (cyan) region of $C_{9'\mu}^{\textsc{np}}$ versus
$C_{9\mu}^{\textsc{np}}$ from the global analysis of $b\to s\mu^+\mu^-$ data performed in
Ref.\,\,\cite{Capdevila:2017bsm}.} \label{c9'c9np}
\end{figure}

Global analyses~\cite{Capdevila:2017bsm,Altmannshofer:2017yso} have found that some of
the best fits to the most recent anomalous $b\to s\mu^+\mu^-$ measurements result from
effective interactions given by
\begin{align} \label{Lb2smm}
{\mathcal L}_{\rm eff}^{} &\,\supset\, \frac{\alpha_{\rm e}^{}\lambda_t^{}G_{\rm F}^{}}
{\sqrt2\,\pi} \big( C_{9\mu}^{}\, \overline{s}_{\,}\gamma^\kappa P_L^{} b + C_{9'\mu}^{} \,
\overline{s}_{\,}\gamma^\kappa P_R^{} b \big) \overline{\mu}_{\,}\gamma_\kappa^{}\mu
\;+\; {\rm H.c.} \,,
\nonumber \\
\lambda_t^{} &\,=\, V_{ts}^*V_{tb}^{} \,, ~~~~~~~
P_{L,R}^{} \,=\, \tfrac{1}{2} (1\mp\gamma_5^{}) \,,
\end{align}
where \,$C_{9\mu}^{}=C_{9\ell}^{\textsc{sm}}+C_{9\mu}^{\textsc{np}}$ and
\,$C_{9'\mu}=C_{9'\mu}^{\textsc{np}}$ are the Wilson coefficients,
\,$\alpha_{\rm e}^{}=1/133$\, is the fine structure constant at the $b$-quark mass $(m_b)$
scale, and $G_{\rm F}$ is the Fermi constant.
The same SM part $C_{9\ell}^{\textsc{sm}}$ occurs in the electron and tau
channels \,$b\to s(e^+e^-,\tau^+\tau^-)$,\, but they are not affected by the NP.
In Fig.\,\ref{c9'c9np}, for later use, we display the 2$\sigma$ (cyan) region of
$C_{9'\mu}^{\textsc{np}}$ versus $C_{9\mu}^{\textsc{np}}$ permitted by the data,
from the global fit carried out in Ref.\,\,\cite{Capdevila:2017bsm}.

\begin{figure}[b]
\includegraphics[width=5in]{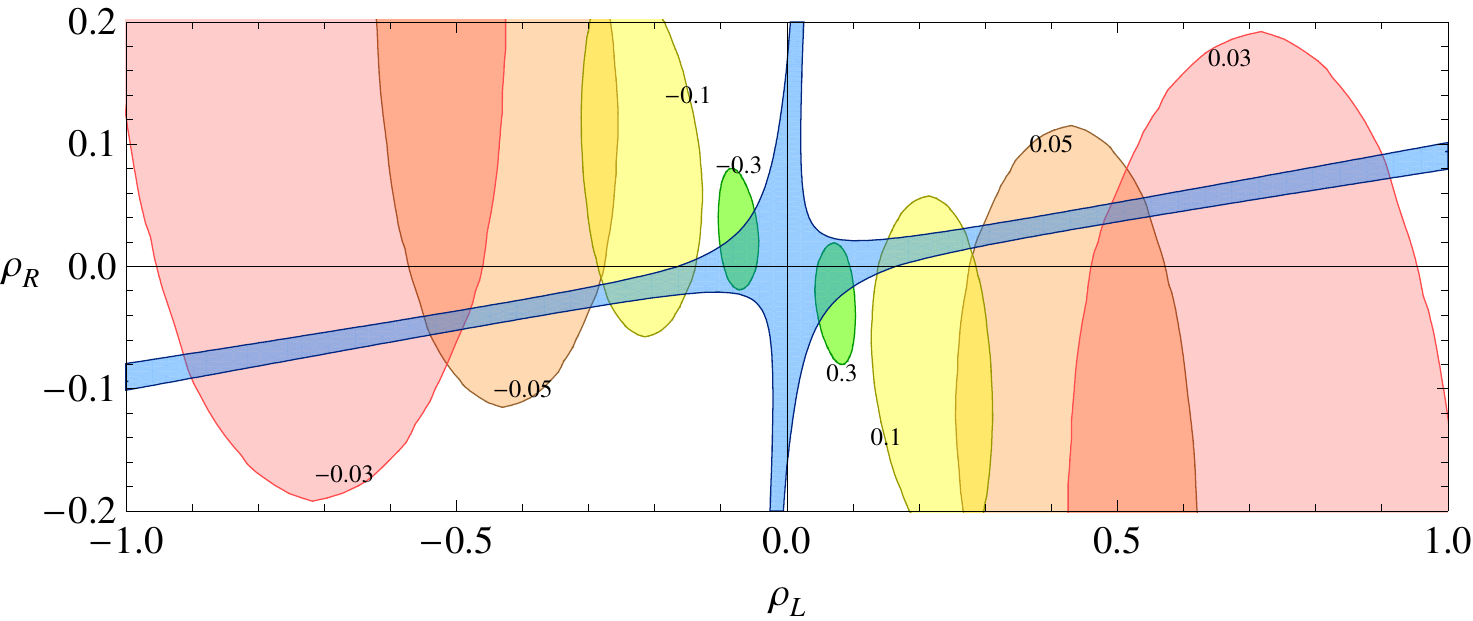}\vspace{-3pt}
\caption{Regions of $\rho_R^{}$ versus $\rho_L^{}$ for $m_{Z'}^{}=1$\,TeV which are consistent
with the $C_{9'\mu}^{\textsc{np}}$-$C_{9\mu}^{\textsc{np}}$ constraint depicted in
Fig.\,\ref{c9'c9np} for \,$\Delta_V^{\mu\mu}=\pm0.03$ (red), $\pm0.05$ (orange),
$\pm0.1$ (yellow), and $\pm0.3$ (green).
The blue area fulfills the condition in Eq.\,(\ref{CBs}) from $B_s$-$\bar B_s$ mixing data.}
\label{roLroR} \end{figure}

In the literature, many models possessing some kind of $Z'$ particle with different sets of
fermionic couplings have been studied in relation to the $b\to s\mu^+\mu^-$
anomalies~\cite{Crivellin:2015mga,Crivellin:2015lwa,Crivellin:2015era,Belanger:2015nma,
Allanach:2015gkd,Fuyuto:2015gmk,Chiang:2016qov,Becirevic:2016zri,Kim:2016bdu,
Altmannshofer:2016jzy,Bhattacharya:2016mcc,Crivellin:2016ejn,Ko:2017quv,Ko:2017lzd,
Kamenik:2017tnu,DiChiara:2017cjq,Ghosh:2017ber,Alok:2017jaf,Alok:2017sui,Wang:2017mrd,
Greljo:2017vvb,Alonso:2017bff,Bonilla:2017lsq,Ellis:2017nrp,Bishara:2017pje,Tang:2017gkz,
Datta:2017ezo,Matsuzaki:2017bpp,DiLuzio:2017chi,Chiang:2017hlj,King:2017anf,Chivukula:2017qsi,
Cline:2017ihf,Chen:2017usq,Baek:2017sew,Bian:2017rpg,Dalchenko:2017shg,Beaudry:2017gtw}.
In the $Z'$ scenario considered here, the interactions responsible for
$C_{9\mu,9'\mu}^{\textsc{np}}$ in Eq.\,\eqref{Lb2smm} are described by
\begin{equation} \label{LsbZ'}
{\mathcal L}_{Z'}^{} \,\supset\,
-\big[ \overline{s}\,\gamma^\kappa\big(\Delta_L^{sb}P_L^{}+\Delta_R^{sb}P_R^{}\big)b\,Z_\kappa'
\;+\; {\rm H.c.} \big] \,-\, \Delta_V^{\mu\mu}\,\overline{\mu}\,\gamma^\kappa\mu\,Z_\kappa' \,,
\end{equation}
where the constants $\Delta_{L,R}^{sb}$ are generally complex and $\Delta_V^{\mu\mu}$ is
real due to the Hermiticity of ${\mathcal L}_{Z'}$.
Any other possible $Z'$ couplings to leptons are taken to be negligible.
To simplify the analysis, hereafter we focus on the special case in which
\begin{equation} \label{Dsb}
\Delta_L^{sb} \;=\; \rho_L^{} V_{ts}^*V_{tb}^{} \,, ~~~~~~~
\Delta_R^{sb} \;=\; \rho_R^{} V_{ts}^*V_{tb}^{} \,,
\end{equation}
where $\rho_{L,R}^{}$ are real numbers, and so they do not supply any new $CP$-violation phase.
Accordingly, for a heavy $Z'$ with mass $m_{Z'}^{}$ we obtain
\begin{equation}
C_{9\mu}^{\textsc{np}} \,=\, \frac{-\sqrt2\,\pi\,\rho_L^{}\Delta_V^{\mu\mu}}
{\alpha_{\rm e}^{}G_{\rm F}^{}\,m_{Z'}^2} \,, ~~~~~~~
C_{9'\mu}^{\textsc{np}} \,=\, \frac{-\sqrt2\,\pi\,\rho_R^{}\Delta_V^{\mu\mu}}
{\alpha_{\rm e}^{}G_{\rm F}^{}\,m_{Z'}^2}  \,.
\end{equation}
In Fig.\,\ref{roLroR}, we illustrate the ranges of $\rho_L^{}$ and $\rho_R^{}$
corresponding to the allowed $C_{9'\mu}^{\textsc{np}}$-$C_{9\mu}^{\textsc{np}}$
(cyan) region in Fig.\,\ref{c9'c9np} for \,$m_{Z'}^{}=1$\,TeV\, and some sample
choices of $\Delta_V^{\mu\mu}$, namely  \,$\pm0.03$ (red), $\pm0.05$ (orange),
$\pm0.1$ (yellow), and $\pm0.3$ (green).
We note that these $\Delta_V^{\mu\mu}$ values contribute positively to the SM muon
anomalous magnetic moment, but with \,$m_{Z'}^{}=1$\,TeV\, are too small to explain
the disagreement with its measurement~\cite{Chiang:2011cv}.

The $Z'$ couplings in Eq.\,(\ref{LsbZ'}) also induce tree-level
effects on \,$\Delta M_s=2|M_{12}^s|$, which pertains to
$B_s$-$\bar B_s$ mixing and has been measured to be \,$\Delta
M_s^{\rm exp}=(17.757\pm0.021)/$ps \cite{pdg}. We can express the
sum of the SM and $Z'$ contributions as~\cite{Buras:2012jb}
\begin{equation} \label{m12s}
M_{12}^s \,=\, M_{12}^{s,\textsc{sm}}  \Bigg( 1 + 4\tilde r\, \frac{\rho_L^2 + \rho_R^2
+ \kappa_{LR}^{}\,\rho_L^{}\rho_R^{}}{g_{\textsc{sm}\,}^2 S_0^{}\,m_{Z'}^2} \Bigg) \,,
\end{equation}
where \cite{Buras:2012jb} \,$\tilde r=0.985$\, for \,$m_{Z'}^{}=1$\,TeV\, is a QCD factor,
\,$g_{\textsc{sm}}^2=1.7814\times10^{-7}\,\rm GeV^{-2}$,\, the SM loop function
\,$S_0^{}\simeq2.35$\, for a top-quark mass \,$m_t^{}\simeq165$\,GeV,\, and
\begin{equation}
\kappa_{LR}^{} \,=\, \frac{6_{\,}\big(C_1^{LR}\big\langle Q_1^{LR}\big\rangle
+ C_2^{LR}\big\langle Q_2^{LR}\big\rangle\big)}
{\eta_B^{}\hat B_{B_s}^{}f_{B_s}^2m_{B_s}^{} \tilde r} \,,
\end{equation}
with \cite{Buras:2012jb}
\,$C_1^{LR}=1-\alpha_{\rm s}^{}\big[1/6+2\log(m_{Z'}/\mu')\big]/(4\pi)$\, and
\,$C_2^{LR}=\alpha_{\rm s}^{}\bigl[-1-12\log(m_{Z'}/\mu')\bigr]/(4\pi)$\, containing the strong
coupling constant $\alpha_{\rm s}^{}$, all evaluated at a scale \,$\mu'\sim m_{Z'}^{}$,\,
\,$\big\langle Q_1^{LR}\big\rangle=-0.37{\rm\,GeV}^3$,\,
\,$\big\langle Q_2^{LR}\big\rangle=0.51{\rm\,GeV}^3$,\, $\eta_B^{}=0.55\pm0.01$,\, and
\cite{utfit} $f_{B_s}^{}\hat B_{B_s}^{1/2}=(262.2\pm9.7){\rm\,MeV}$.\,
With the central values of these parameters and $m_{B_s}$ from Ref.\,\cite{pdg},
we get \,$\kappa_{LR}^{}=-11.2$\, for \,$m_{Z'}^{}=1$\,TeV.

To apply restrictions on $\rho_{L,R}^{}$ from the $B_s$-$\bar B_s$ mixing data, we impose
\begin{equation} \label{CBs}
0.899 \,\le\, \frac{\Delta M_s^{}}{\Delta M_s^{\textsc{sm}}}
=\bigg|\frac{M_{12}^s}{M_{12}^{s,\textsc{sm}}}\bigg| \,\le\, 1.252 \,,
\end{equation}
which is the 95\% confidence level (CL) range from the latest UTfit global analysis~\cite{utfit}.
Since some of the numbers quoted in the last paragraph have uncertainties up to a few
percent, we let $\kappa_{LR}^{}$ vary by up to 10\% from its central value when scanning
the parameter space for $\rho_{L,R}^{}$ values which conform to Eq.\,(\ref{CBs}).
For \,$m_{Z'}^{}=1$\,TeV,\, we incorporate the scan result into Fig.\,\ref{roLroR},
represented by the blue area.
Thus, in this figure each overlap of the blue area with one of the other colored ones of
a particular $\Delta_V^{\mu\mu}$ value corresponds to the parameter space that can explain
the $b\to s\mu^+\mu^-$ anomalies and simultaneously satisfies Eq.\,(\ref{CBs}).
With smaller choices of $|\Delta_V^{\mu\mu}|$, such overlaps could be found at larger
$|\rho_{L,R}^{}|$ values.
This graph also reveals that in the absence of the right-handed coupling, \,$\rho_R^{}=0$,\,
the allowed range of $\rho_L^{}$ would be rather narrow, indicating the importance of
nonvanishing $\rho_R^{}$ for gaining bigger viable parameter space~\cite{Crivellin:2015era}.

\section{\boldmath$Z'$ contributions to rare nonleptonic $\bar B_s$ decays\label{Bs2MM'}}

Given that \,$\bar B_s\to(\eta,\eta',\phi)(\pi^0,\rho^0)$\, change both strangeness and isospin,
in the SM their amplitudes proceed from \,$b\to s$\, four-quark operators $O_{1,2}^u$ and
$O_{7,8,9,10}$ which are derived from charmless tree and electroweak-penguin diagrams, respectively.
In contrast, the QCD-penguin operators $O_{3,4,5,6}$, which preserve isospin symmetry, do not
affect these processes.\footnote{The expressions for $O_i$, $i=1,2,\cdots,10$, can be found in,
{\it e.g.}, \cite{Williamson:2006hb}.}
In many models beyond the SM, new interactions may modify the Wilson coefficients $C_i$ of $O_i$
and/or give rise to extra operators $\tilde O_i$ which are the chirality-flipped counterparts
of $O_i$.
A flavor-violating $Z'$ boson may contribute to some of them, depending on the details of its
properties.

In our scenario of interest, besides its couplings in Eq.\,(\ref{LsbZ'}), the $Z'$ has
flavor-conserving interactions with the $u$ and $d$ quarks via
\begin{equation} \label{LqqZ'}
{\mathcal L}_{Z'}^{} \,\supset\,
-\big[ \overline{u}\,\gamma^\kappa\big(\Delta_L^{uu}P_L^{}+\Delta_R^{uu}P_R^{}\big)u +
\overline{d}\,\gamma^\kappa\big(\Delta_L^{dd}P_L^{}+\Delta_R^{dd}P_R^{}\big)d \big]Z_\kappa' \,,
\end{equation}
the constants $\Delta_{L,R}^{uu,dd}$ being real, but does not couple flavor-diagonally to other
quarks.
From Eqs.\,\,\eqref{LsbZ'} and \eqref{LqqZ'}, we can derive tree-level $Z'$-mediated diagrams
contributing to nonleptonic \,$b\to s$\, reactions.
For a heavy $Z'$, these diagrams yield
\begin{equation} \label{bsqq}
{\cal L}_{4\mbox{\scriptsize-quark}}^{Z'} \,\supset\, \frac{-\lambda_t}{m_{Z'}^2}\,
\overline{s}\,\gamma^\kappa\big(\rho_L^{}P_L^{}+\rho_R^{}P_R^{}\big)b~
\raisebox{3pt}{\footnotesize$\displaystyle\sum_{q=u,d}$}\, \overline{q}\,\gamma_\kappa^{}
\big(\Delta_L^{qq}P_L^{}+\Delta_R^{qq}P_R^{}\big)q
\end{equation}
after applying Eq.\,(\ref{Dsb}).
It is straightforward to realize that these additional terms bring about modifications to
the coefficients of the QCD- and electroweak-penguin operators $O_{3,5,7,9}$ in the SM and
also generate their chirality-flipped partners $\tilde O_{3,5,7,9}$~\cite{Barger:2009qs}.
We can express them in the effective Lagrangian for \,$b\to s$\, transitions as
\begin{align}
{\cal L}_{\rm eff}^{} \,\supset&~ \sqrt8\,\lambda_t^{~}G_{\rm F}^{}
\raisebox{3pt}{\footnotesize$\displaystyle\sum_{q=u,d}$}
\Big\{ \overline{s}\,\gamma^\kappa P_L^{}b
\Big[ \Big( C_3^{}+\tfrac{3}{2}C_9^{}e_q^{} \Big) \overline{q}\,\gamma_\kappa^{}P_L^{} q
+ \Big( C_5^{}+\tfrac{3}{2}C_7^{}e_q^{} \Big) \overline{q}\,\gamma_\kappa^{}P_R^{} q \Big]
\nonumber \\ & \hspace{1in} +\,
\overline{s}\,\gamma^\kappa P_R^{}b \Big[ \Big( \tilde C_3^{}
+ \tfrac{3}{2}\tilde C_9^{}e_q^{} \Big) \overline{q}\,\gamma_\kappa^{}P_R^{} q
+ \Big( \tilde C_5^{}+\tfrac{3}{2}\tilde C_7^{}e_q^{} \Big)
\overline{q}\,\gamma_\kappa^{}P_L^{} q \Big] \Big\} \,,
\end{align}
where \,$C_j^{}=C_j^{\textsc{sm}} + C_j^{Z'}$\, and \,$\tilde C_j^{}=\tilde C_j^{Z'}$\, for
\,$j=3,5,7,9$\, are the Wilson coefficients.
Thus, from Eq.\,\eqref{bsqq} we have~\cite{Hofer:2010ee,Chang:2013hba,Barger:2009qs}
\begin{align} \label{cz'}
C_{3,5}^{Z'} &\,=\, \frac{\rho_L^{}\bigl(-\Delta_{L,R}^{uu}-2\Delta_{L,R}^{dd}\big)}
{6\sqrt2\,G_{\rm F}^{}\,m_{Z'}^2}
\,=\, \frac{\rho_L^{}\bigl(-\delta_{L,R}^{}-3\Delta_{L,R}^{dd}\big)}
{6\sqrt2\,G_{\rm F}^{}\,m_{Z'}^2} \,, ~~~~~~~
\nonumber \\
\tilde C_{3,5}^{Z'} &\,=\, \frac{\rho_R^{}\bigl(-\Delta_{R,L}^{uu}-2\Delta_{R,L}^{dd}\big)}
{6\sqrt2\,G_{\rm F}^{}\,m_{Z'}^2}
\,=\, \frac{\rho_R^{}\bigl(-\delta_{R,L}^{}-3\Delta_{R,L}^{dd}\big)}
{6\sqrt2\,G_{\rm F}^{}\,m_{Z'}^2} \,,
\nonumber \\
C_{7,9}^{Z'} &\,=\, \frac{\rho_L^{}\bigl(-\Delta_{R,L}^{uu}+\Delta_{R,L}^{dd}\big)}
{3\sqrt2\,G_{\rm F}^{}\,m_{Z'}^2}
\,=\, \frac{-\rho_{L\,}^{}\delta_{R,L}^{}}{3\sqrt2\,G_{\rm F}^{}\,m_{Z'}^2} \,, ~~~~~~~~~
\nonumber \\
\tilde C_{7,9}^{Z'} &\,=\, \frac{\rho_R^{}\bigl(-\Delta_{L,R}^{uu}+\Delta_{L,R}^{dd}\big)}
{3\sqrt2\,G_{\rm F}^{}\,m_{Z'}^2}
\,=\, \frac{-\rho_{R\,}^{}\delta_{L,R}^{}}{3\sqrt2\,G_{\rm F}^{}\,m_{Z'}^2}  \,,
\end{align}
where
\begin{equation} \label{dLdR}
\delta_L^{} \,=\, \Delta_L^{uu} - \Delta_L^{dd} \,, ~~~~~~~
\delta_R^{} \,=\, \Delta_R^{uu} - \Delta_R^{dd} \,.
\end{equation}
As $O_{3,5}$ and $\tilde O_{3,5}$ do not break isospin, only $C_{7,9}^{Z'}$ and
$\tilde C_{7,9}^{Z'}$ contribute to \,$\bar B_s\to(\eta,\eta',\phi)(\pi^0,\rho^0)$.\,

To estimate the $Z'$ impact on these decays, we make use of the soft-collinear effective theory
(SCET)\,\,\cite{Bauer:2000ew,Bauer:2000yr,Chay:2003zp,Bauer:2004tj,Bauer:2005kd,
Williamson:2006hb,Wang:2008rk}, similarly to what was done in Ref.\,\cite{Faisel:2014dna} in
the case of a leptophobic-$Z'$ model.
For any one of them, the SCET amplitude at leading order in $\alpha_s(m_b)$ can be written
as \cite{Wang:2008rk}
\begin{align} \label{AB2MM'}
{\cal A}_{\bar B_s\to M_1 M_2}^{} \,=&~\, \frac{f_{M_1}^{}G_{\rm F}^{}m_{B_s}^2}{\sqrt2} \bigg[
\int_0^1 d\nu\, \Big( \zeta_J^{BM_2\,} T_{1J}^{}(\nu) + \zeta_{Jg}^{BM_2\,} T_{1Jg}^{}(\nu) \Big)
\phi_{M_1}^{}(\nu)
+ \zeta^{BM_2\,} T_1^{} + \zeta_g^{BM_2\,} T_{1g}^{} \bigg]
\nonumber \\ & +\, (1\leftrightarrow 2) \;,
\end{align}
where $f_M^{}$ is the decay constant of meson $M$, the $\zeta$'s are nonperturbative hadronic
parameters which can be fixed from experiment, the $T$'s are hard kernels which are functions of
the Wilson coefficients $C_i$ and $\tilde C_i$, and $\phi_{M}^{}(\nu)$ is the light-cone
distribution amplitude of $M$ which is normalized as \,$\int_0^1d\nu\,\phi_M(\nu)=1$.\,
The so-called charming-penguin term, which in this case conserves isospin, is absent from
${\cal A}_{\bar B_s\to M_1 M_2}$.
The hard kernels for the decays of concern are available from the
literature\,\,\cite{Wang:2017rmh,Williamson:2006hb,Wang:2008rk} and have been listed in
Table\,\,\ref{kernels}, where the flavor states
\,$\eta_q\sim\big(u\bar u+d\bar d\big)/\sqrt2$\, and
\,$\eta_s^{}\sim s\bar s$\, are related to the physical meson states $\eta$ and $\eta'$ by
\,$\eta=\eta_q\cos\theta-\eta_s^{}\sin\theta$\, and
\,$\eta'=\eta_q\sin\theta+\eta_s^{}\cos\theta$\, with mixing angle
\,$\theta=39.3^\circ$
\,\cite{Williamson:2006hb,Wang:2008rk,Feldmann:1998vh}.

\begin{table}[t]
\begin{tabular}{|c||c|c|c|c|} \hline
\,Decay mode\, & $~T_1~$ & $T_2$ & $~T_{1g}~$ & $T_{2g}\vphantom{\int_|^|}$ \\
\hline\hline
$\bar{B}_s\to\eta_s^{}\pi^0$ & 0 & ~$\frac{1}{\sqrt2} (c_2-c_3)$~ & 0 &
~$\frac{1}{\sqrt2} (c_2-c_3)\vphantom{\int_{\int_|}^{\int}}$~  \\
$\bar{B}_s\to\eta_s^{}\rho^0$ & 0 & $\frac{1}{\sqrt2} (c_2+c_3)$ & 0 &
$\frac{1}{\sqrt2} (c_2+c_3)\vphantom{\int_{\int_|}^{\int}}$ \\
$\bar{B}_s\to\eta_q\pi^0$ & 0 & 0 & 0 & $c_2-c_3\vphantom{\int_{\int_|}^{\int}}$  \\
$\bar{B}_s\to\eta_q\rho^0$ & 0 & 0 & 0 & $c_2+c_3\vphantom{\int_{\int_|}^{\int}}$ \\
$\bar{B}_s\to\phi\pi^0$ & 0 &
$\frac{1}{\sqrt2} (c_2-c_3)\vphantom{\int_{\int_|}^{\int}}$ & 0 & 0 \\
$\bar{B}_s\to\phi\rho^0  $ & 0 &
$\frac{1}{\sqrt2} (c_2+c_3)\vphantom{\int_{\int_|}^{\int}}$ & 0 & 0 \\ \hline
\end{tabular} \smallskip
\caption{Hard kernels $T_{1,2,1g,2g}^{}$ for \,$\bar B_s\to(\eta,\eta',\phi)(\pi^0,\rho^0)$\,
decays.
The hard kernels $T_{rJ,rJg}^{}(\nu)$ for \,$r=1,2$\, are obtainable from $T_{r,rg}^{}$,
respectively, through the replacement \,$c_k^{}\to b_k^{}$,\, where $b_k^{}$ has dependence
on $\nu$.} \label{kernels}
\end{table}

In the presence of NP which also generates the extra operators $\tilde O_i$, the quantities
$c_{2,3}$ and $b_{2,3}$ in Table\,\,\ref{kernels} depend not only on $C_i$
and $\tilde C_i$, but also on the final mesons $M_1$ and $M_2$, as well as on the CKM factors
$\lambda_t$ and \,$\lambda_u=V_{us}^*V_{ub}^{}$.\,
The dependence on $M_1$ and $M_2$ arises from the fact that, with regard to the nonzero
kernels in this table, for each 4-quark operator the $\bar B_s$\,$\to$\,$M_1$ and
vacuum\,$\to$\,$M_2$ matrix elements and their contraction in the amplitude can lead to
an overall negative or positive sign for the contribution of the operator, the sign being
determined by the chirality combination of the operator and by whether the final mesons are
pseudoscalars ($PP$), vectors ($VV$), $PV$, or $VP$.
Thus, for \,$\bar{B}_s\to(\eta_q,\eta_s^{})\pi^0$\, and \,$\bar{B}_s\to\phi\rho^0$\,  we
have\footnote{The formula for $b_2$ given in \cite{Faisel:2014dna} contains typos which we have
corrected here in Eq.\,\eqref{cb}.}
\begin{align} \label{cb}
c_2^{} &\,=\, \lambda_u^{} \Bigg( C_{2}-\tilde{C}_2+\frac{C_{1}-\tilde{C}_1}{N_{\rm c}} \Bigg)
- \frac{3\lambda_t}{2} \Bigg(C_{9}-\tilde{C}_9+\frac{C_{10}-\tilde{C}_{10}}{N_{\rm c}}\Bigg) \,,
\nonumber \\
c_3^{} &\,=\, -\frac{3\lambda_t}{2} \Bigg(C_7-\tilde{C}_7+\frac{C_8-\tilde{C}_8}{N_{\rm c}}\Bigg) \,,
\nonumber \\
b_2^{} &\,=\, \lambda_u^{} \Bigg[ C_{2}-\tilde{C}_2 + \bigg(1-\frac{m_b^{}}{\omega_3}\bigg)
\frac{C_{1}-\tilde{C}_1}{N_{\rm c}} \Bigg] - \frac{3 \lambda_t}{2} \Bigg[ C_{9}-\tilde{C}_9
+ \bigg(1-\frac{m_b^{}}{\omega_3}\bigg) \frac{C_{10}-\tilde{C}_{10}}{N_{\rm c}} \Bigg] \,,
\nonumber \\
b_3^{} &\,=\, -\frac{3 \lambda_t}{2} \Bigg[ C_7-\tilde{C}_7 + \bigg(1-\frac{m_b^{}}{\omega_2}\bigg)
\frac{C_8-\tilde{C}_8}{N_{\rm c}} \Bigg] \,,
\end{align}
where \,$N_{\rm c}=3$\, is the color number and $b_{2,3}$, which are contained in $T_{2J}(\nu)$
and $T_{2Jg}(\nu)$, are also functions of $\nu$ via \cite{Wang:2008rk}
\,$\omega_2^{}=\nu m_{B_s}$\, and \,$\omega_3^{}=(\nu-1)m_{B_s}$.\,
However, for \,$\bar{B}_s\to(\eta_q,\eta_s^{})\rho^0$\, and \,$\bar{B}_s\to\phi\pi^0$\, we need
to make the sign change \,$-\tilde C_i\to+\tilde C_i$\, in $c_{2,3}$ and $b_{2,3}$.

The expressions in Eq.\,\eqref{cb} generalize the SM ones provided previously in
Refs.\,\,\cite{Williamson:2006hb,Wang:2008rk}.
They also supplied the values of the SM
coefficients $C_i^{\textsc{sm}}$ at the $m_b$ scale, \,$C_{1,2}^{\textsc{sm}}=(1.11,-0.253)$\,
and \,$C_{7,8,9,10}^{\textsc{sm}}=(0.09,0.24,-10.3,2.2)\times10^{-3}$ \cite{Buchalla:1995vs},
which we will use in $c_{2,3}$ and $b_{2,3}$.
Our $Z'$ contributions of interest, in Eq.\,(\ref{cz'}), enter Eq.\,\eqref{cb} only via
\,$C_{7,9}^{}=C_{7,9}^{\textsc{sm}}+C_{7,9}^{Z'}$\, and
\,$\tilde C_{7,9}^{}=\tilde C_{7,9}^{Z'}$.\,

For numerical computation of ${\cal A}_{\bar B_s\to M_1 M_2}$, in view of Table\,\,\ref{kernels},
the meson decay constants which we need are only \,$f_\pi^{}=131$\,MeV\, and
\,$f_\rho^{}=209$\,MeV,\, and the integral in Eq.\,(\ref{AB2MM'}) can be treated
with the aid of the relations
\,$\int_0^1d\nu\,\phi_M^{}(\nu)/\nu=\int_0^1d\nu\,\phi_M^{}(\nu)/(1-\nu)\equiv
\langle\chi^{-1}\rangle_M^{}$\,
for \,$M=\pi,\rho$,\, in which cases \,$\langle\chi^{-1}\rangle_\pi^{}=3.3$\, and
\,$\langle\chi^{-1}\rangle_\rho^{}=3.45$\, \cite{Williamson:2006hb,Wang:2008rk}.
Moreover, for the $\zeta$'s we adopt the two solutions derived from the fit to data done in
Ref.\,\cite{Wang:2008rk}:
\begin{align} \label{zetas}
\big(\zeta^P,\zeta_J^P,\zeta^V,\zeta_J^V,\zeta_g^{},\zeta_{Jg}^{}\big)_1 &\,=\,
(0.137,0.069,0.117,0.116,-0.049,-0.027) \,, \nonumber \\
\big(\zeta^P,\zeta_J^P,\zeta^V,\zeta_J^V,\zeta_g^{},\zeta_{Jg}^{}\big)_2 &\,=\,
(0.141,0.056,0.227,0.065,-0.100,0.051) \,.
\end{align}
From these, we can obtain \,$\zeta_{(J)}^{B\eta_q}=\zeta_{(J)}^{B\eta_s}=\zeta_{(J)}^P$,
\,$\zeta_{(J)}^{B\phi}=\zeta_{(J)}^V$,\, and
\,$\zeta_{(J)g}^{B\eta_q}=\zeta_{(J)g}^{B\eta_s}=\zeta_{(J)g}$\, under the assumption of
flavor-SU(3) symmetry\,\,\cite{Wang:2008rk}.
In Eq.\,\eqref{zetas}, we have not displayed the errors of the $\zeta$s from the fit to
data, which are available from Ref.\,\cite{Wang:2008rk}.
Other input parameters that we will employ are the meson masses
\,$m_{\pi^0}=134.977$, \,$m_{\eta}=547.862$, \,$m_{\eta'}=957.78$, \,$m_{\rho^0}=769$,
\,$m_\phi=1019.46$,\, and \,$m_{B_s}=5366.89$,\, all in units of MeV, and the $B_s$ lifetime
\,$\tau_{B_s}^{}=1.505\times10^{-12}$\,s,\, which are their central values from
Ref.\,\cite{pdg}.

\begin{table}[b] \smallskip
\begin{tabular}{|c||c|c|c|c|} \hline
\multirow{2}{*}{$\begin{array}{c}\rm Decay \vspace{-3pt} \\ \rm mode \end{array}$} &
\multirow{2}{*}{QCDF} & \multirow{2}{*}{PQCD} & \multicolumn{2}{c|}{SCET} \\
\cline{4-5} & & & Solution 1 & Solution 2 \\ \hline\hline
$\bar B_s\to\eta\pi^0$    & \,$0.05_{-0.01-0.01}^{+0.03+0.02}$\, &
\,$0.05 _{-0.02-0.01-0.00}^{+0.02+0.01+0.00}$\, & \,$0.032\pm0.013\pm0.008$\, &
\,$0.025\pm0.010\pm0.003\vphantom{|_{|_o}^{\int^o}}$\,
\\
\,$\bar B_s\to\eta'\pi^0$\, & $0.04_{-0.00-0.00}^{+0.01+0.01}$ &
$0.11_{-0.03-0.01-0.00}^{+0.05+0.02+0.00}$ & $0.001\pm0.000\pm0.005$ &
$0.052\pm0.021\pm0.015\vphantom{|_{|_o}^{\int^o}}$
\\
$\bar B_s\to\phi\pi^0$ & $0.12_{-0.01-0.02}^{+0.02+0.04}$ & $0.16_{-0.05-0.02-0}^{+0.06+0.02+0}$
& \,$0.074\pm0.030\pm0.009$\, & $0.091\pm0.036\pm0.016\vphantom{|_{|_o}^{\int^o}}$
\\ \hline
$\bar B_s\to\eta\rho^0$ & $0.10_{-0.01-0.01}^{+0.02+0.02}$ &
$0.06_{-0.02-0.01-0.00}^{+0.03+0.01+0.00}$ & $0.078\pm0.031\pm0.022$ &
\,$0.059\pm0.023\pm0.006\vphantom{|_{|_o}^{\int^o|}}$\,
\\
$\bar B_s\to\eta'\rho^0$  & $0.16_{-0.02-0.03}^{+0.06+0.03}$ &
$0.13_{-0.04-0.02-0.01}^{+0.06+0.02+0.00}$ & $0.003\pm0.001\pm0.013$ &
$0.141\pm0.056\pm0.042\vphantom{|_{|_o}^{\int^o}}$
\\ \hline\hline
$\bar B_s\to\phi\rho^0$  & \,$0.18_{-0.01-0.04}^{+0.01+0.09}$\, &
\,$0.23_{-0.07-0.01-0.01}^{+0.09+0.03+0.00}$\, &
\multicolumn{2}{c|}{$0.36\pm0.14\pm0.04\vphantom{|_{|_o}^{\int^o}}$}
\\ \hline
\end{tabular}
\caption{Branching fractions, in units of $10^{-6}$, of
\,$\bar B_s\to(\eta,\eta',\phi)(\pi^0,\rho^0)$\, decays in the SM.
For the first five modes, the last two columns correspond to the two solutions of SCET
parameters in Eq.\,(\ref{zetas}).
The errors of the SCET predictions are due to assumed 20\% flavor-SU(3)-breaking effects
and the errors in the $\zeta$s from fits to data, respectively.
For comparison, the second and third columns contain results calculated in the frameworks
of QCDF \cite{Cheng:2009mu} and PQCD \cite{Ali:2007ff}.} \label{smbf} \vspace{-1ex}
\end{table}

Before addressing the $Z'$ influence on \,$\bar B_s\to(\eta,\eta',\phi)(\pi^0,\rho^0)$,\,
we provide the SM predictions for their branching fractions, which are collected in
Table\,\,\ref{smbf}.
For the first five modes, the SCET numbers have been evaluated with the preceding formulas and
parameter values, and the last two columns correspond to the two solutions of SCET parameters
in Eq.\,(\ref{zetas}).
For the sixth $(\phi\rho^0)$ mode, the SCET entry has been computed with the CKM and SCET
parameters supplied very recently in Ref.\,\,\cite{Wang:2017rmh}.
The two errors in each of the SCET predictions are due to flavor-SU(3)-breaking effects
which we have assumed to be 20\% and due to the errors in the $\zeta$s from the fits to data,
respectively, the latter errors being given in Refs.\,\cite{Wang:2017rmh,Wang:2008rk}.
The SCET numbers for \,$\bar B_s\to(\eta,\eta')\rho^0,\phi\pi^0$
$\big(\bar B_s\to\phi\rho^0\big)$ are close to the corresponding ones determined in
Ref.\,\cite{Wang:2008rk}~(\cite{Wang:2017rmh}).\footnote{The SCET predictions in
Table\,\,\ref{smbf} differ from those obtained in \cite{Faisel:2014dna} because some of the input
parameters used in our two papers are not the same.}
For comparison, in the second and third columns we quote numbers
calculated with QCD factorization (QCDF) \cite{Cheng:2009mu} and perturbative QCD (PQCD)
\cite{Ali:2007ff}.
Evidently, these two methods produce results comparable to those of SCET, especially with its
Solution 2 in the case of the first five modes, considering the errors in the predictions.
The entries for \,$\bar B_s\to\phi\rho^0$ are also compatible with the new
measurement
\,${\cal B}\big(\bar B_s\to\phi\rho^0\big){}_{\rm exp}^{}=(0.27\pm0.08)\!\times\!10^{-6}$\,
\cite{pdg} mentioned earlier.
An important implication of what we see in this table is that NP would not be easily
noticeable in the rates of these decays unless it could enhance them by more than a factor of 2.
This possibility may be unlikely to be realized in the case of $\bar B_s\to\phi\rho^0$ which has
been detected having a rate consistent with SM expectations.
Nevertheless, as we demonstrate below, substantial enhancement can still occur in some of
the other channels.

Now we include the $Z'$ contributions from Eq.\,(\ref{cz'}) in order to examine their impact on
these decays.
As Table\,\,\ref{smbf} indicates that the predictions of the SCET Solution 1 for
\,$\bar B_s\to\eta'\big(\pi^0,\rho^0\big)$\, are comparatively quite suppressed, from this
point on we employ only Solution 2 parameters in our treatment of
\,$\bar B_s\to(\eta,\eta')(\pi^0,\rho^0),\phi\pi^0$.\,
Thus, summing the SM and $Z'$ terms for \,$m_{Z'}^{}=1$\,TeV, with the central values of
the input parameters, we find the amplitudes (in units of GeV) for the $\pi^0$ channels to be
\begin{align} \label{pimodes}
10^9\,{\cal A}_{\bar{B}_s\to\eta\pi^0} &\,\simeq\,
 1.67 + 0.47 i + (3.96 - 0.08 i) (\rho_L^{}+\rho_R^{})(\delta_L-\delta_R) \,, \nonumber \\
10^9\,{\cal A}_{\bar{B}_s\to\eta'\pi^0} &\,\simeq\,
 0.48 - 2.48 i - (1.90 - 0.04 i) (\rho_L^{}+\rho_R^{})(\delta_L-\delta_R) \,, \nonumber \\
10^9\,{\cal A}_{\bar{B}_s\to\phi\pi^0} &\,\simeq\,
-2.88 - 1.69 i - (7.85 - 0.15 i) (\rho_L^{}-\rho_R^{})(\delta_L-\delta_R)
\end{align}
and for the $\rho^0$ channels
\begin{align} \label{rhomodes}
10^9\,{\cal A}_{\bar{B}_s\to\eta\rho^0} &\,\simeq\,
 2.56 + 0.77 i + (6.32 - 0.12 i) (\rho_L^{}+\rho_R^{})(\delta_L+\delta_R) \,, \nonumber \\
10^9\,{\cal A}_{\bar{B}_s\to\eta'\rho^0} &\,\simeq\,
 0.78 - 4.12 i - (3.03 - 0.06 i) (\rho_L^{}+\rho_R^{})(\delta_L+\delta_R) \,, \nonumber \\
10^9\,{\cal A}_{\bar{B}_s\to\phi\rho^0} &\,\simeq\,
-6.53 - 1.47 i - (15.3 - 0.3 i) (\rho_L^{}-\rho_R^{})(\delta_L+\delta_R) \,,
\end{align}
where $\delta_{L,R}$ are defined in Eq.\,\eqref{dLdR}.

\begin{figure}[b] \smallskip
\includegraphics[width=9cm]{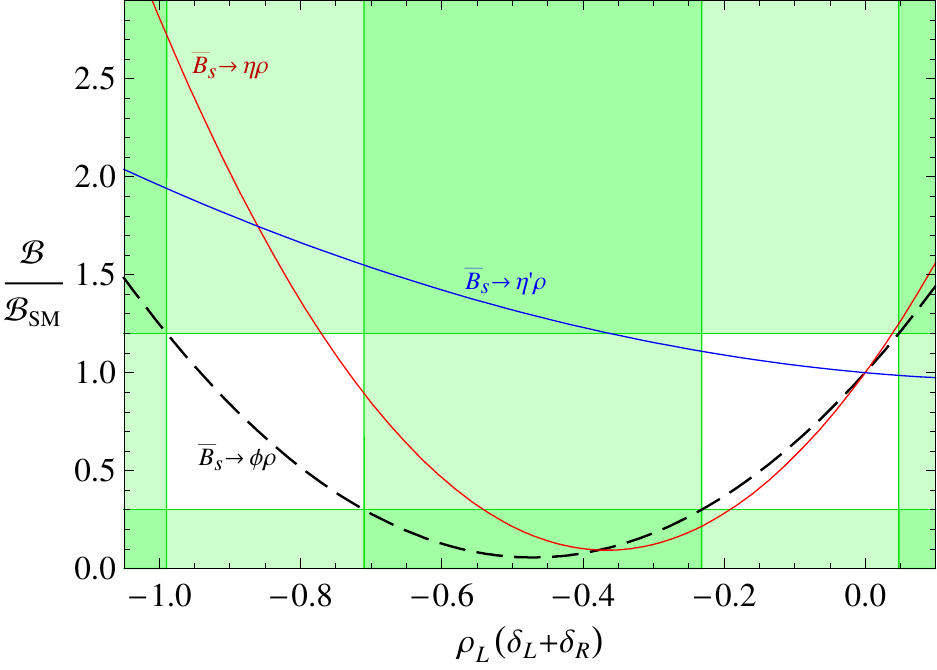}\vspace{-7pt}
\caption{The calculated branching fractions of \,$\bar B_s\to\eta\rho^0$ (red solid curve),
\,$\bar B_s\to\eta'\rho^0$ (blue solid curve), and \,$\bar B_s\to\phi\rho^0$ (black curve),
normalized by their respective SM predictions listed in Table\,\,\ref{smbf},
versus the product \,$\rho_L^{}(\delta_L+\delta_R)$\, in the case where
\,$\rho_R^{}=0.1\,\rho_L^{}$\, and \,$m_{Z'}^{}=1$\,TeV.\,
The vertical length of the unshaded areas and the \,$\rho_L^{}(\delta_L+\delta_R)$\, values
within them satisfy the restriction in Eq.\,\eqref{constr1}.} \label{B-rhomodes} \vspace{-1ex}
\end{figure}

We notice that the amplitudes in Eqs.\,\,\eqref{pimodes} and \eqref{rhomodes} do not all have
the same dependence on $\rho_{L,R}^{}$ and $\delta_{L,R}$.
Therefore, although ${\cal B}\big(\bar B_s\to\phi\rho^0\big){}_{\rm exp}^{}$ implies a restraint
on the values of \,$(\rho_L^{}-\rho_R^{})(\delta_L+\delta_R)$\, in
${\cal A}_{\bar{B}_s\to \phi\rho^0}$, the amplitudes for the other channels, which have
different combinations of $\rho_{L,R}^{}$ and $\delta_{L,R}$, may generally still be altered
considerably with respect to their SM parts.
However, in our particular $Z'$ case $\rho_{L,R}^{}$ must satisfy
\,$\rho_R^{}\sim0.1\,\rho_L^{}$,\, as can be inferred from Fig.\,\ref{roLroR}.
Hence, based on Eq.\,(\ref{rhomodes}), we may expect that the amplitudes for
\,$\bar{B}_s\to(\eta,\eta')\rho^0$\, do not deviate hugely from their SM values.
To look into this more concretely, for definiteness we take \,$\rho_R^{}=0.1\,\rho_L^{}$\,
and impose
\begin{equation} \label{constr1}
0.11 \,\le\, 10^6\,{\cal B}\big(\bar B_s\to\phi\rho^0\big) \,\le\, 0.43 \,,
\end{equation}
which is the 2$\sigma$ range of ${\cal B}\big(\bar B_s\to\phi\rho^0\big){}_{\rm exp}^{}$.
From the allowed values of the product \,$\rho_L^{}(\delta_L+\delta_R)$\, we can
assess how much the branching fractions of \,$\bar B_s\to(\eta,\eta')\rho^0$\, are
modified compared to the central values of their respective SM predictions in
Table\,\,\ref{smbf} under SCET Solution 2.
We show the results in Fig.\,\ref{B-rhomodes}, which also depicts Eq.\,\eqref{constr1}
relative to the SM prediction.
We further find that the ranges \,$\rho_L^{}(\delta_L+\delta_R)\in[-0.99,-0.71]$ and
$[-0.23,0.048]$\, fulfill Eq.\,\eqref{constr1}.
Within these ranges, represented by the horizontal portions of the unshaded areas in
this figure, we learn that ${\cal B}\big(\bar B_s\to\eta\rho^0\big)$ (red
solid curve) can reach up to {\footnotesize$\sim\,$}2.7 times its SM value, whereas for
\,$\bar B_s\to\eta'\rho^0$\, (blue solid curve) the enhancement is at most about 1.9
times.\footnote{Before the LHCb detection of the \,$\bar B_s\to\phi\rho^0$\, evidence
\cite{Aaij:2016qnm}, the possibilities of the \,$\bar B_s\to(\phi,\eta)\rho^0$\,
rates exceeding their SM predictions by an order of magnitude were entertained in
\cite{Hofer:2010ee,Faisel:2014dna}, respectively, for the \,$\delta_L^{}=0$\, case.}

\begin{figure}[t]
\includegraphics[width=4in]{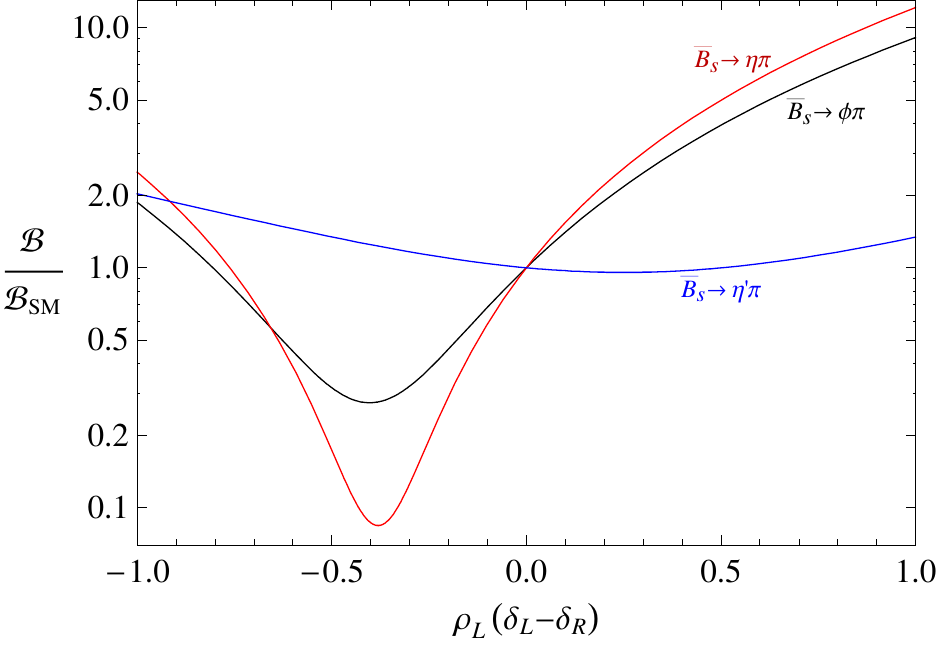}\vspace{-5pt}
\caption{The calculated branching fractions of \,$\bar B_s\to\eta\pi^0$ (red curve),
\,$\bar B_s\to\eta'\pi^0$ (blue curve), and \,$\bar B_s\to\phi\pi^0$ (black curve),
normalized by their respective SM values listed in Table\,\,\ref{smbf} under SCET Solution 2,
versus the product \,$\rho_L^{}(\delta_L-\delta_R)$\, in the case where
\,$\rho_R^{}=0.1\,\rho_L^{}$\, and \,$m_{Z'}^{}=1$\,TeV.} \label{B-pimodes}
\end{figure}

For \,$\bar B_s\to(\eta,\eta',\phi)\pi^0$,\, the $Z'$-induced terms in Eq.\,\eqref{pimodes} are
proportional to \,$\delta_L-\delta_R$.\,
Therefore, these channels are not subject to the condition in Eq.\,\eqref{constr1}, and their
amplitudes may be affected by the $Z'$ contributions more than their $\rho^0$ counterparts.
To examine this more quantitatively, we set \,$\rho_R^{}=0.1\,\rho_L^{}$\, as in the previous
paragraph and subsequently compute the branching fractions of these
$\pi^0$ modes for \,$-1\le\rho_L^{}(\delta_L-\delta_R)\le1$.\,
In Fig.\,\ref{B-pimodes} we present the results divided by the central values of their
respective SM predictions in Table\,\,\ref{smbf} under SCET Solution 2.
We observe that over most of the \,$\rho_L^{}(\delta_L-\delta_R)>0$\, region covered in this plot
the $Z'$ effects can cause the branching fractions of \,$\bar B_s\to(\eta,\phi)\pi^0$ to exceed
their SM counterparts by at least a factor of 2 and up to about an order of magnitude.
Moreover, the \,$\bar B_s\to(\eta,\phi)\pi^0$\, rates tend to be enhanced together with roughly
similar enlargement factors.
One also notices that the $Z'$ impact could instead bring about substantial reduction of their rates.
In Table\,\,\ref{B/Bexp}, we provide examples of the enhancement factors for some representative
values of \,$\rho_L^{}(\delta_L-\delta_R)$.\,

\begin{table}[t] \bigskip
\begin{tabular}{|c||c|c|c|} \hline
~$\rho_L^{}(\delta_L-\delta_R)$~ & ~$\bar B_s\to\eta\pi^0$~ & ~$\bar B_s\to\eta'\pi^0$~ &
~$\bar B_s\to\phi\pi^0\vphantom{\int_{|_|}^{\int^|}}$~ \\ \hline\hline
$   -1$ & 2.5 & 2.0 & 1.9$\vphantom{\int_|^{|_|}}$  \\
$-0.5$ & 0.17 & 1.3 & 0.32$\vphantom{\int_|^|}$ \\
$ 0.5$ & 5.0 & 1.0 & 3.9$\vphantom{\int_|^|}$  \\
$    1$ & 12 & 1.3 & 9.1$\vphantom{\int_|^|}$ \\ \hline
\end{tabular} \smallskip
\caption{Enhancement factors of the branching fractions of
\,$\bar B_s\to(\eta,\eta',\phi)\pi^0$\, with respect to their SM predictions at a few
representative values of \,$\rho_L^{}(\delta_L-\delta_R)$ in the
\,$\rho_R^{}=0.1\,\rho_L^{}$\, and \,$m_{Z'}^{}=1$\,TeV\, case.} \label{B/Bexp}
\end{table}

It is worth remarking that the $Z'$-generated coefficients $C_{3,5,7,9}^{Z'}$ and
$\tilde C_{3,5,7,9}^{Z'}$ in Eq.\,\eqref{cz'} enter the amplitudes for nonleptonic \,$b\to s$\,
decays which are not dominated by the contributions of the electroweak-penguin operators,
such as \,$B_s\to K^{(*)}\bar K^{(*)}$\, and \,$B\to\pi K^{(*)}$.\,
Since these transitions have been observed, their data imply restrictions on the size of
\,$\rho_L^{}(\delta_L\pm\delta_R)$,\, as the requirement \,$\rho_R^{}\sim0.1\,\rho_L^{}$\, in
our $Z'$ scenario implies that the role of $\tilde C_{3,5,7,9}^{Z'}$ is minor.
Our choices \,$|\rho_L^{}(\delta_L\pm\delta_R)|\le1$\, above correspond to
$\big|C_7^{Z'}\pm C_9^{Z'}\big|\le0.0202\simeq2|C_9^{\textsc{sm}}|$,\, where
\,$C_9^{\textsc{sm}}=-0.0103$\, as quoted before.
We have checked that the changes to the rates of those decays due to
\,$|C_{7,9}^{Z'}|\lesssim|C_9^{\textsc{sm}}|$\, are less than the uncertainties of
the SCET estimates in the SM, which are typically around 20\% to
40\%~\cite{Wang:2017rmh,Williamson:2006hb,Wang:2008rk}.
As for the influence of $C_{3,5}^{Z'}$, it can be minimized by adjusting the extra free
parameters $\Delta_{L,R}^{dd}$ in Eq.\,\eqref{cz'}.\footnote{For instance, selecting
\,$2\Delta_{L,R}^{dd}=-\Delta_{L,R}^{uu}$\, would lead to
\,$C_{3,5}^{Z'}=0$,\, which was considered in \cite{Hofer:2010ee,Faisel:2014dna}.}
For comparison, earlier studies~\cite{Hofer:2010ee,Chang:2013hba,Bobeth:2014rra} concerning
potential NP in \,$\bar B_s\to\phi(\pi^0,\rho^0)$\, pointed out that rate enhancement
factors of a few to an order of magnitude could still occur in the $\phi\pi^0$ mode and
that \,$|C_j^{\textsc{np}}/C_9^{\textsc{sm}}|\lesssim\rm 2$\, was not yet disfavored.

Finally, we would like to mention that the $Z'$ coupling parameters of interest are separately
consistent with constraints which may be pertinent from collider measurements.
We illustrate this with the specific examples in Table\,\,\ref{couplings} for different
sets of \,$\rho_L^{}(\delta_L\pm\delta_R)$ and $\Delta_V^{\mu\mu}$ values in the aforesaid
case where \,$\rho_R^{}=0.1\,\rho_L^{}$\, and \,$m_{Z'}^{}=1$\,TeV.\,
The choice \,$\big(\rho_L^{},\Delta_V^{\mu\mu}\big)=(0.8,0.03)$\, is evidently within
the region covered in Fig.\,\ref{roLroR}, while the points
\,$\big(\rho_L^{},\Delta_V^{\mu\mu}\big)=\big((1.0,1.2),0.02\big)$\, lie in
the extension thereof.
In this table, the displayed numbers for $\delta_{L,R}$ can comfortably comply with
the condition
\,$|\delta_{L,R}|\le1.0\big[1+(1.3{\rm\,TeV})^2/m_{Z'}^2\big]m_{Z'}^{}$/(3\,TeV)\,
inferred in Ref.\,\cite{Buras:2015jaq} from the study on LHC bounds
in Ref.\,\cite{deVries:2014apa}.
For the lepton sector, the results of Refs.\,\,\cite{Chiang:2017hlj,Chiang:2013aha} imply that
the selected $\Delta_V^{\mu\mu}$ values are compatible with LEP data on $Z$-boson decays into
lepton pairs~\cite{pdg}.
Furthermore, as the $\bar e e Z'$ interaction is supposed to be vanishing, restraints
from LEP II measurements on \,$e^+e^-\to f\bar f$\, can be evaded.
Lastly, LHC searches for new high-mass phenomena in the dilepton final states have
the potential for significantly probing $\delta_{L,R}$ and $\Delta_V^{\mu\mu}$ at the same time.
Nevertheless, their samples values in Table\,\,\ref{couplings} can be checked to be consistent
with the most recent \,$pp\to\ell^+\ell^-X$\, results from the ATLAS
experiment\,\,\cite{Aaboud:2017buh}.\footnote{We may test our $Z'$ coupling choices with
the latest ATLAS\,\,\cite{Aaboud:2017buh} constraint on a nonstandard quark-muon contact
interaction of the form
\,${\cal L}=(4\pi/\Lambda^2)\eta_{\chi\chi'}\,\bar q_\chi\gamma^\beta q_\chi\,
\bar\mu_{\chi'}\gamma_\beta^{}\mu_{\chi'}$,\,
where $\Lambda$ is a heavy mass scale, \,$\eta_{\chi\chi'}=-1(1)$\, if the new and SM contributions
to \,$q\bar q\to\mu^+\mu^-$\, interfere constructively (destructively), and \,$\chi,\chi'=L,R$.\,
It turns out that the strongest restriction applies to the \,$\chi\chi'=RL$ or $RR$\, case and
arises from the 95\%-CL limit \,$\Lambda>28$\,TeV \,\cite{Aaboud:2017buh} corresponding to
\,$4\pi/\Lambda^2<0.016{\rm\;TeV}^{-2}$.\,
This can be fulfilled by \,$\big|\Delta_R^{qq}\Delta_V^{\mu\mu}\big|$ for \,$q=u,d$\,
and the entries in the last two columns of Table\,\,\ref{couplings} with selections such as
\,$\Delta_R^{uu}=-\Delta_R^{dd}=\delta_R^{}/2$.}

\begin{table}[h] \bigskip
\begin{tabular}{|c|c||c|c|c||c|} \hline
~$\rho_L^{}(\delta_L+\delta_R)$~ & ~$\rho_L^{}(\delta_L-\delta_R)$~ & ~$\rho_L^{}$~ &
~$\delta_L$~ & ~$\delta_R$~ & ~$\Delta_V^{\mu\mu}\vphantom{\int_|^\int}$~ \\ \hline\hline
$-0.85$ & 0.5 & ~0.8~ & ~$-0.219$~ & ~$-0.844$~ & ~0.03$\vphantom{\int_|^\int}$~ \\
$-0.90$ & 0.7 & 1.0 &  $-0.1 $ & $-0.8$  &  0.02$\vphantom{\int_|^|}$  \\
$-0.95$ & 0.9 & 1.2 &  $-0.021$ & $-0.771$ & ~0.02$\vphantom{\int_|^|}$~ \\
$-0.99$ & 1.0 & 1.2 & ~0.004 & ~$-0.829$~ &  0.02$\vphantom{\int_|^|}$  \\ \hline
\end{tabular} \smallskip
\caption{The quark-$Z'$ coupling parameters $\rho_L^{}$ and $\delta_{L,R}$\, corresponding to
a few sample sets of \,$\rho_L^{}(\delta_L\pm\delta_R)$ and $\Delta_V^{\mu\mu}$ in
the \,$\rho_R^{}=0.1\,\rho_L^{}$\, and \,$m_{Z'}^{}=1$\,TeV\, case.} \label{couplings}
\end{table}

\section{Conclusions\label{conclusions}}

We have explored the possibility that the recently observed anomalies in several
\,$b\to s\mu^+\mu^-$\, processes are attributable to the interactions of a $Z'$ boson which
also contribute to rare nonleptonic decays of the $\bar B_s$ meson,
namely $\bar B_s\to(\eta,\eta',\phi)(\pi^0,\rho^0)$.
Given that the amplitudes for these purely isospin-violating decays have CKM-suppressed
tree components and tend to be controlled mainly by the electroweak-penguin operators,
their decay rates are expected to be relatively small in the SM, making these modes
potentially sensitive to signals beyond the SM.
The $Z'$ couplings are subject to various restrictions, particularly from the data on
$B_s$-$\bar B_s$ mixing and the new experimental finding on $\bar B_s^0\to\phi\rho^0$,
besides the measurements of \,$b\to s\mu^+\mu^-$\, transitions.
We showed that, within the allowed parameter space, the $Z'$ impact on
$\bar B_s^0\to(\eta,\phi)\pi^0$ can cause their rates to grow up to an order of magnitude greater
than their expectations in the SM.
On the other hand, the enhancement factors  for $\bar B_s^0\to\eta'\pi^0,(\eta,\eta')\rho^0$ are
at most a few.
The different enlargement factors of these different channels depend not only on the combinations
of the $Z'$ couplings occurring in their amplitudes, but also on how the SM and $Z'$ terms in
the amplitudes interfere with each other.
Therefore, the observations of more of these decays in future experiments, together with
improved upcoming data on \,$b\to s\mu^+\mu^-$,\, will test our $Z'$ model more comprehensively.

\acknowledgements

The work of J.T was supported in part by the Republic of China Ministry of Education
Academic Excellence Program (Grant No. 105R891505).
He would like to thank Seungwon Baek and Pyungwon Ko for generous hospitality at the Korea
Institute for Advanced Study where parts of this research were done.


\begin{thebibliography}{0}

\bibitem{Aaij:2015oid}
  R.~Aaij {\it et al.} [LHCb Collaboration],
  JHEP {\bf 1602}, 104 (2016)  [arXiv:1512.04442 [hep-ex]].

\bibitem{Wehle:2016yoi}
  S.~Wehle {\it et al.} [Belle Collaboration],
  Phys.\ Rev.\ Lett.\  {\bf 118}, no. 11, 111801 (2017)  [arXiv:1612.05014 [hep-ex]].

\bibitem{Aaij:2014ora}
  R.~Aaij {\it et al.} [LHCb Collaboration],
  Phys.\ Rev.\ Lett.\  {\bf 113}, 151601 (2014)  [arXiv:1406.6482 [hep-ex]].

\bibitem{Aaij:2017vbb}
  R.~Aaij {\it et al.} [LHCb Collaboration],
  JHEP {\bf 1708}, 055 (2017)  [arXiv:1705.05802 [hep-ex]].

\bibitem{Hiller:2003js}
  G.~Hiller and F.~Kruger,
  Phys.\ Rev.\ D {\bf 69}, 074020 (2004)  [hep-ph/0310219].

\bibitem{Bouchard:2013mia}
  C.~Bouchard {\it et al.} [HPQCD Collaboration],
  Phys.\ Rev.\ Lett.\  {\bf 111}, no. 16, 162002 (2013)
  Erratum: [Phys.\ Rev.\ Lett.\  {\bf 112}, no. 14, 149902 (2014)]  [arXiv:1306.0434 [hep-ph]].

\bibitem{Bordone:2016gaq}
  M.~Bordone, G.~Isidori, and A.~Pattori,
  Eur.\ Phys.\ J.\ C {\bf 76}, no. 8, 440 (2016)  [arXiv:1605.07633 [hep-ph]].

\bibitem{Aaij:2014pli}
  R.~Aaij {\it et al.} [LHCb Collaboration],
  JHEP {\bf 1406}, 133 (2014)  [arXiv:1403.8044 [hep-ex]].

\bibitem{Aaij:2015esa}
  R.~Aaij {\it et al.} [LHCb Collaboration],
  JHEP {\bf 1509}, 179 (2015)  [arXiv:1506.08777 [hep-ex]].

\bibitem{pdg}
  C.~Patrignani {\it et al.} [Particle Data Group],
  Chin.\ Phys.\ C {\bf 40}, no. 10, 100001 (2016)
and 2017 update.

\bibitem{Descotes-Genon:2013wba}
  S.~Descotes-Genon, J.~Matias, and J.~Virto,
  Phys.\ Rev.\ D {\bf 88}, 074002 (2013)  [arXiv:1307.5683 [hep-ph]].

\bibitem{Beaujean:2013soa}
  F.~Beaujean, C.~Bobeth, and D.~van Dyk,
  Eur.\ Phys.\ J.\ C {\bf 74}, 2897 (2014); 3179(E) (2014)  [arXiv:1310.2478 [hep-ph]].

\bibitem{Hurth:2013ssa}
  T.~Hurth and F.~Mahmoudi,
  JHEP {\bf 1404}, 097 (2014)  [arXiv:1312.5267 [hep-ph]].

\bibitem{Alonso:2014csa}
  R.~Alonso, B.~Grinstein, and J.~Martin Camalich,
  Phys.\ Rev.\ Lett.\  {\bf 113}, 241802 (2014)  [arXiv:1407.7044 [hep-ph]].

\bibitem{Hiller:2014yaa}
  G.~Hiller and M.~Schmaltz,
  Phys.\ Rev.\ D {\bf 90}, 054014 (2014)  [arXiv:1408.1627 [hep-ph]].

\bibitem{Ghosh:2014awa}
  D.~Ghosh, M.~Nardecchia, and S.A.~Renner,
  JHEP {\bf 1412}, 131 (2014)  [arXiv:1408.4097 [hep-ph]].

\bibitem{Hurth:2014vma}
  T.~Hurth, F.~Mahmoudi, and S.~Neshatpour,
  JHEP {\bf 1412}, 053 (2014)  [arXiv:1410.4545 [hep-ph]].

\bibitem{Capdevila:2017bsm}
  B.~Capdevila, A.~Crivellin, S.~Descotes-Genon, J.~Matias, and J.~Virto,
  arXiv:1704.05340 [hep-ph].

\bibitem{Altmannshofer:2017yso}
  W.~Altmannshofer, P.~Stangl, and D.M.~Straub,
  Phys.\ Rev.\ D {\bf 96}, no. 5, 055008 (2017)  [arXiv:1704.05435 [hep-ph]].

\bibitem{DAmico:2017mtc}
  G.~D'Amico, M.~Nardecchia, P.~Panci, F.~Sannino, A.~Strumia, R.~Torre, and A.~Urbano,
  JHEP {\bf 1709}, 010 (2017)  [arXiv:1704.05438 [hep-ph]].

\bibitem{Hiller:2017bzc}
  G.~Hiller and I.~Nisandzic,
  Phys.\ Rev.\ D {\bf 96}, no. 3, 035003 (2017)  [arXiv:1704.05444 [hep-ph]].

\bibitem{Geng:2017svp}
  L.S.~Geng, B.~Grinstein, S.~J\"ager, J.~Martin Camalich, X.L.~Ren, and R.X.~Shi,
Phys.\ Rev.\ D {\bf 96}, no. 9, 093006 (2017)  [arXiv:1704.05446 [hep-ph]].

\bibitem{Ciuchini:2017mik}
  M.~Ciuchini, A.M.~Coutinho, M.~Fedele, E.~Franco, A.~Paul, L.~Silvestrini, and M.~Valli,
  Eur.\ Phys.\ J.\ C {\bf 77}, no. 10, 688 (2017)  [arXiv:1704.05447 [hep-ph]].

\bibitem{Celis:2017doq}
  A.~Celis, J.~Fuentes-Martin, A.~Vicente, and J.~Virto,
 Phys.\ Rev.\ D {\bf 96}, no. 3, 035026 (2017)  [arXiv:1704.05672 [hep-ph]].

\bibitem{Hurth:2017hxg}
  T.~Hurth, F.~Mahmoudi, D.~Martinez Santos, and S.~Neshatpour,
  Phys.\ Rev.\ D {\bf 96}, no. 9, 095034 (2017)  [arXiv:1705.06274 [hep-ph]].

\bibitem{Fleischer:1994rs}
  R.~Fleischer,
  Phys.\ Lett.\ B {\bf 332}, 419 (1994).

\bibitem{Deshpande:1994yd}
  N.G.~Deshpande, X.G.~He, and J.~Trampetic,
  Phys.\ Lett.\ B {\bf 345}, 547 (1995)  [hep-ph/9410388].

\bibitem{Tseng:1998wm}
  B.~Tseng,
  Phys.\ Lett.\ B {\bf 446}, 125 (1999)  [hep-ph/9807393];
  Y.H.~Chen, H.Y.~Cheng, and B.~Tseng,
  Phys.\ Rev.\ D {\bf 59}, 074003 (1999)  [hep-ph/9809364].

\bibitem{Aaij:2016qnm}
  R.~Aaij {\it et al.} [LHCb Collaboration],
  Phys.\ Rev.\ D {\bf 95}, no. 1, 012006 (2017)  [arXiv:1610.05187 [hep-ex]].

\bibitem{Cheng:2009mu}
  H.Y.~Cheng and C.K.~Chua,
  Phys.\ Rev.\ D {\bf 80}, 114026 (2009)  [arXiv:0910.5237 [hep-ph]].

\bibitem{Ali:2007ff}
  A.~Ali, G.~Kramer, Y.~Li, C.D.~Lu, Y.L.~Shen, W.~Wang, and Y.M.~Wang,
  Phys.\ Rev.\ D {\bf 76}, 074018 (2007)  [hep-ph/0703162].

\bibitem{Wang:2017rmh}
  C.~Wang, S.H.~Zhou, Y.~Li, and C.D.~Lu,
  Phys.\ Rev.\ D {\bf 96}, no. 7, 073004 (2017)  [arXiv:1708.04861 [hep-ph]].

\bibitem{Faisel:2011kq}
  G.~Faisel,
  JHEP {\bf 1208}, 031 (2012)  [arXiv:1106.4651 [hep-ph]];
  Phys.\ Lett.\ B {\bf 731}, 279 (2014)  [arXiv:1311.0740 [hep-ph]].

\bibitem{Hofer:2010ee}
  L.~Hofer, D.~Scherer, and L.~Vernazza,
  JHEP {\bf 1102}, 080 (2011)  [arXiv:1011.6319 [hep-ph]].

\bibitem{Hua:2010wf}
  J.~Hua, C.S.~Kim, and Y.~Li,
  Phys.\ Lett.\ B {\bf 690}, 508 (2010)  [arXiv:1002.2532 [hep-ph]].

\bibitem{Chang:2013hba}
  Q.~Chang, X.Q.~Li, and Y.D.~Yang,
  J.\ Phys.\ G {\bf 41}, 105002 (2014)  [arXiv:1312.1302 [hep-ph]].

\bibitem{Faisel:2014dna}
  G.~Faisel,
  Eur.\ Phys.\ J.\ C {\bf 77}, no. 6, 380 (2017)  [arXiv:1412.3011 [hep-ph]].

\bibitem{Bobeth:2014rra}
  C.~Bobeth, M.~Gorbahn, and S.~Vickers,
  Eur.\ Phys.\ J.\ C {\bf 75}, no. 7, 340 (2015)  [arXiv:1409.3252 [hep-ph]].

\bibitem{Bauer:2000ew}
  C.W.~Bauer, S.~Fleming, and M.E.~Luke,
  Phys.\ Rev.\  D {\bf 63}, 014006 (2000)  [arXiv:hep-ph/0005275].

\bibitem{Bauer:2000yr}
  C.W.~Bauer, S.~Fleming, D.~Pirjol, and I.W.~Stewart,
  Phys.\ Rev.\  D {\bf 63}, 114020 (2001)  [arXiv:hep-ph/0011336].

\bibitem{Chay:2003zp}
  J.~Chay and C.~Kim,
  Phys.\ Rev.\ D {\bf 68}, 071502 (2003)  [arXiv:hep-ph/0301055];
  Nucl.\ Phys.\ B {\bf 680}, 302 (2004)  [arXiv:hep-ph/0301262].

\bibitem{Bauer:2004tj}
  C.W.~Bauer, D.~Pirjol, I.Z.~Rothstein, and I.W.~Stewart,
  Phys.\ Rev.\ D {\bf 70}, 054015 (2004)  [hep-ph/0401188].

\bibitem{Bauer:2005kd}
  C.W.~Bauer, I.Z.~Rothstein, and I.W.~Stewart,
  Phys.\ Rev.\ D {\bf 74}, 034010 (2006)  [hep-ph/0510241].

\bibitem{Williamson:2006hb}
  A.R.~Williamson and J.~Zupan,
  Phys.\ Rev.\  D {\bf 74}, 014003 (2006)  [Erratum-ibid.\  D {\bf 74}, 03901 (2006)]
  [arXiv:hep-ph/0601214].

\bibitem{Wang:2008rk}
W.~Wang, Y.M.~Wang, D.S.~Yang, and C.D.~Lu,
  Phys.\ Rev.\  D {\bf 78}, 034011 (2008)  [arXiv:0801.3123 [hep-ph]].


\bibitem{Crivellin:2015mga}
  A.~Crivellin, G.~D'Ambrosio and J.~Heeck,
  Phys.\ Rev.\ Lett.\  {\bf 114}, 151801 (2015)  [arXiv:1501.00993 [hep-ph]].

\bibitem{Crivellin:2015lwa}
  A.~Crivellin, G.~D'Ambrosio, and J.~Heeck,
  Phys.\ Rev.\ D {\bf 91}, no. 7, 075006 (2015)  [arXiv:1503.03477 [hep-ph]].

\bibitem{Crivellin:2015era}
  A.~Crivellin, L.~Hofer, J.~Matias, U.~Nierste, S.~Pokorski and J.~Rosiek,
  Phys.\ Rev.\ D {\bf 92}, no. 5, 054013 (2015)  [arXiv:1504.07928 [hep-ph]].

\bibitem{Belanger:2015nma}
  G.~Belanger, C.~Delaunay, and S.~Westhoff,
  Phys.\ Rev.\ D {\bf 92}, 055021 (2015)  [arXiv:1507.06660 [hep-ph]].

\bibitem{Allanach:2015gkd}
  B.~Allanach, F.S.~Queiroz, A.~Strumia, and S.~Sun,
  Phys.\ Rev.\ D {\bf 93}, no. 5, 055045 (2016)  [arXiv:1511.07447 [hep-ph]].

\bibitem{Fuyuto:2015gmk}
  K.~Fuyuto, W.S.~Hou, and M.~Kohda,
  Phys.\ Rev.\ D {\bf 93}, no. 5, 054021 (2016)  [arXiv:1512.09026 [hep-ph]].

\bibitem{Chiang:2016qov}
  C.W.~Chiang, X.G.~He, and G.~Valencia,
  Phys.\ Rev.\ D {\bf 93}, no. 7, 074003 (2016)  [arXiv:1601.07328 [hep-ph]].

\bibitem{Becirevic:2016zri}
  D.~Becirevic, O.~Sumensari, and R.~Zukanovich Funchal,
  Eur.\ Phys.\ J.\ C {\bf 76}, no. 3, 134 (2016)  [arXiv:1602.00881 [hep-ph]].

\bibitem{Kim:2016bdu}
  C.S.~Kim, X.B.~Yuan, and Y.J.~Zheng,
  Phys.\ Rev.\ D {\bf 93}, no. 9, 095009 (2016)  [arXiv:1602.08107 [hep-ph]].

\bibitem{Altmannshofer:2016jzy}
  W.~Altmannshofer, S.~Gori, S.~Profumo, and F.S.~Queiroz,
  JHEP {\bf 1612}, 106 (2016)  [arXiv:1609.04026 [hep-ph]].

\bibitem{Bhattacharya:2016mcc}
  B.~Bhattacharya, A.~Datta, J.P.~Guvin, D.~London, and R.~Watanabe,
  JHEP {\bf 1701}, 015 (2017)  [arXiv:1609.09078 [hep-ph]].

\bibitem{Crivellin:2016ejn}
  A.~Crivellin, J.~Fuentes-Martin, A.~Greljo, and G.~Isidori,
  Phys.\ Lett.\ B {\bf 766}, 77 (2017)  [arXiv:1611.02703 [hep-ph]].

\bibitem{Ko:2017quv}
  P.~Ko, T.~Nomura and H.~Okada,
  Phys.\ Lett.\ B {\bf 772}, 547 (2017)  [arXiv:1701.05788 [hep-ph]];
  Phys.\ Rev.\ D {\bf 95}, no. 11, 111701 (2017)  [arXiv:1702.02699 [hep-ph]].

\bibitem{Ko:2017lzd}
  P.~Ko, Y.~Omura, Y.~Shigekami, and C.~Yu,
  Phys.\ Rev.\ D {\bf 95}, no. 11, 115040 (2017)  [arXiv:1702.08666 [hep-ph]].

\bibitem{Kamenik:2017tnu}
  J.F.~Kamenik, Y.~Soreq, and J.~Zupan,
  arXiv:1704.06005 [hep-ph].

\bibitem{DiChiara:2017cjq}
  S.~Di Chiara, A.~Fowlie, S.~Fraser, C.~Marzo, L.~Marzola, M.~Raidal, and C.~Spethmann,
 Nucl.\ Phys.\ B {\bf 923}, 245 (2017)  [arXiv:1704.06200 [hep-ph]].

\bibitem{Ghosh:2017ber}
  D.~Ghosh,
  Eur.\ Phys.\ J.\ C {\bf 77}, no. 10, 694 (2017)  [arXiv:1704.06240 [hep-ph]].

\bibitem{Alok:2017jaf}
  A.K.~Alok, D.~Kumar, J.~Kumar, and R.~Sharma,
  arXiv:1704.07347 [hep-ph].

\bibitem{Alok:2017sui}
  A.K.~Alok, B.~Bhattacharya, A.~Datta, D.~Kumar, J.~Kumar, and D.~London,
  Phys.\ Rev.\ D {\bf 96}, no. 9, 095009 (2017)  [arXiv:1704.07397 [hep-ph]].

\bibitem{Wang:2017mrd}
  W.~Wang and S.~Zhao,
  Chin.\ Phys.\ C {\bf 42}, no. 1, 013105 (2018)  [arXiv:1704.08168 [hep-ph]].

\bibitem{Greljo:2017vvb}
  A.~Greljo and D.~Marzocca,
  Eur.\ Phys.\ J.\ C {\bf 77}, no. 8, 548 (2017)  [arXiv:1704.09015 [hep-ph]].

\bibitem{Alonso:2017bff}
  R.~Alonso, P.~Cox, C.~Han, and T.T.~Yanagida,
  Phys.\ Rev.\ D {\bf 96}, no. 7, 071701 (2017)  [arXiv:1704.08158 [hep-ph]].
  Phys.\ Lett.\ B {\bf 774}, 643 (2017)  [arXiv:1705.03858 [hep-ph]].

\bibitem{Bonilla:2017lsq}
  C.~Bonilla, T.~Modak, R.~Srivastava, and J.W.F.~Valle,
  arXiv:1705.00915 [hep-ph].

\bibitem{Ellis:2017nrp}
  J.~Ellis, M.~Fairbairn and P.~Tunney,
  arXiv:1705.03447 [hep-ph].

\bibitem{Bishara:2017pje}
  F.~Bishara, U.~Haisch, and P.F.~Monni,
 Phys.\ Rev.\ D {\bf 96}, no. 5, 055002 (2017)  [arXiv:1705.03465 [hep-ph]].

\bibitem{Tang:2017gkz}
  Y.~Tang and Y.L.~Wu,
  arXiv:1705.05643 [hep-ph];

\bibitem{Datta:2017ezo}
  A.~Datta, J.~Kumar, J.~Liao, and D.~Marfatia,
  arXiv:1705.08423 [hep-ph].

\bibitem{Matsuzaki:2017bpp}
  S.~Matsuzaki, K.~Nishiwaki and R.~Watanabe,
  JHEP {\bf 1708}, 145 (2017)  [arXiv:1706.01463 [hep-ph]].

\bibitem{DiLuzio:2017chi}
  L.~Di Luzio and M.~Nardecchia,
  Eur.\ Phys.\ J.\ C {\bf 77}, no. 8, 536 (2017)  [arXiv:1706.01868 [hep-ph]].

\bibitem{Chiang:2017hlj}
  C.W.~Chiang, X.G.~He, J.~Tandean, and X.B.~Yuan,
  Phys.\ Rev.\ D {\bf 96}, no. 11, 115022 (2017)  [arXiv:1706.02696 [hep-ph]].

\bibitem{King:2017anf}
  S.F.~King,
  JHEP {\bf 1708}, 019 (2017)  [arXiv:1706.06100 [hep-ph]].

\bibitem{Chivukula:2017qsi}
  R.S.~Chivukula, J.~Isaacson, K.A.~Mohan, D.~Sengupta, and E.H.~Simmons,
  Phys.\ Rev.\ D {\bf 96}, no. 7, 075012 (2017)  [arXiv:1706.06575 [hep-ph]].

\bibitem{Cline:2017ihf}
  J.M.~Cline and J.~Martin Camalich,
  Phys.\ Rev.\ D {\bf 96}, no. 5, 055036 (2017)  [arXiv:1706.08510 [hep-ph]].

\bibitem{Chen:2017usq}
  C.H.~Chen and T.~Nomura,
  Phys.\ Lett.\ B {\bf 777}, 420 (2018)  [arXiv:1707.03249 [hep-ph]].

\bibitem{Baek:2017sew}
  S.~Baek,
  arXiv:1707.04573 [hep-ph].

\bibitem{Bian:2017rpg}
  L.~Bian, S.M.~Choi, Y.J.~Kang, and H.M.~Lee,
  Phys.\ Rev.\ D {\bf 96}, no. 7, 075038 (2017)  [arXiv:1707.04811 [hep-ph]].

\bibitem{Dalchenko:2017shg}
  M.~Dalchenko, B.~Dutta, R.~Eusebi, P.~Huang, T.~Kamon, and D.~Rathjens,
  arXiv:1707.07016 [hep-ph].

\bibitem{Beaudry:2017gtw}
  N.B.~Beaudry, A.~Datta, D.~London, A.~Rashed, and J.S.~Roux,
  JHEP {\bf 1801}, 074 (2018)  [arXiv:1709.07142 [hep-ph]].


\bibitem{Chiang:2011cv}
  C.W.~Chiang, Y.F.~Lin, and J.~Tandean,
  JHEP {\bf 1111}, 083 (2011)   [arXiv:1108.3969 [hep-ph]].

\bibitem{Buras:2012jb}
  A.J.~Buras, F.~De Fazio, and J.~Girrbach,
  JHEP {\bf 1302}, 116 (2013)  [arXiv:1211.1896 [hep-ph]].

\bibitem{utfit} 
  M.~Bona {\it et al.} [UTfit Collaboration],
  JHEP {\bf 0803}, 049 (2008)  [arXiv:0707.0636 [hep-ph]].
Online updates available at http://www.utfit.org.

\bibitem{Barger:2009qs}
  V.~Barger, L.~L.~Everett, J.~Jiang, P.~Langacker, T.~Liu and C.~E.~M.~Wagner,
  JHEP {\bf 0912}, 048 (2009)  [arXiv:0906.3745 [hep-ph]].

\bibitem{Buchalla:1995vs}
  G.~Buchalla, A.J.~Buras, and M.E.~Lautenbacher,
  Rev.\ Mod.\ Phys.\  {\bf 68}, 1125 (1996)  [hep-ph/9512380].

\bibitem{Feldmann:1998vh}
  T.~Feldmann, P.~Kroll and B.~Stech,
  Phys.\ Rev.\ D {\bf 58}, 114006 (1998)  [hep-ph/9802409];
  Phys.\ Lett.\ B {\bf 449}, 339 (1999)  [hep-ph/9812269].

\bibitem{Buras:2015jaq}
  A.J.~Buras,
  JHEP {\bf 1604}, 071 (2016)  [arXiv:1601.00005 [hep-ph]].

\bibitem{deVries:2014apa}
  M.~de Vries,
  JHEP {\bf 1503}, 095 (2015)  [arXiv:1409.4657 [hep-ph]].

\bibitem{Aaboud:2017buh}
  M.~Aaboud {\it et al.} [ATLAS Collaboration],
  JHEP {\bf 1710}, 182 (2017)  [arXiv:1707.02424 [hep-ex]].

\bibitem{Chiang:2013aha}
  C.W.~Chiang, T.~Nomura, and J.~Tandean,
  Phys.\ Rev.\ D {\bf 87}, 075020 (2013)  [arXiv:1302.2894 [hep-ph]].

\end{thebibliography}
\end{document}